%% file: npl.tex
\documentclass[aps,prd,showpacs,superscriptaddress,reprint]{revtex4-1}
\usepackage{amsmath,amsfonts,amssymb}
\usepackage{dsfont,graphicx}
\usepackage{bm}
\graphicspath{{Plots/}}
\DeclareMathOperator{\Tr}{Tr}
\newcommand{\bs}{\begin{subequations}}
\newcommand{\es}{\end{subequations}}
\newcommand{\adb}{\allowdisplaybreaks } 
\newcommand{\ann}{\adb \nonumber \\}
\newcommand{\TE}{\mathsf{TE}}
\newcommand{\TM}{\mathsf{TM}}

\newcommand{\cA}{\mathcal{A}}
\newcommand{\cB}{\mathcal{B}}
\newcommand{\cZ}{\mathcal{Z}}
\newcommand{\cN}{\mathcal{N}}
\usepackage{color} \definecolor{darkgreen}{rgb}{0,.5,0}
\usepackage[colorlinks,filecolor=blue,citecolor=darkgreen,unicode]{hyperref}
\begin{document}
\title{The Casimir effect for a stack of conductive planes}

\author{Nail Khusnutdinov}\email{nail.khusnutdinov@gmail.com}
\affiliation{Institute of Physics, Kazan Federal University, Kremlevskaya 18, Kazan, 420008, Russia}
\affiliation{Centro de Matem\'atica, Computa\c{c}\~ao e Cogni\c{c}\~ao, Universidade Federal do ABC, 09210-170 Santo Andr\'e, SP, Brazil}
\author{Rashid Kashapov}\email{kashapov.rashid@gmail.com}
\affiliation{Institute of Physics, Kazan Federal University, Kremlevskaya 18, Kazan, 420008, Russia}
\author{Lilia M. Woods}
\affiliation{Department of Physics, University of South Florida, Tampa, Florida 33620, USA}
\date{\today}
\begin{abstract}
The Casimir interaction in a stack of equally spaced infinitely thin layers is investigated within the zero-frequency mode summation method. The response properties are considered to be described by a constant conductivity or by a Drude-Lorentz model with a  finite set of oscillators consistent with the optical characteristics for graphite. It is found that the asymptotic distance dependence is affected significantly by the specific response. While the energy is $\sim 1/d^3$ for the constant conductivity model, the energy exhibits fractional dependence $\sim 1/d^{5/2}$ for the Drude-Lorentz description. The Casimir force  on a plane is also strongly dependent upon the particular plane location in the stack. Furthermore, the calculated Casimir energy within the Drude-Lorentz model yields results in good agreement with measured cohesion energy in graphite.  
\end{abstract}  
\maketitle 

\section{Introduction}

Dispersive  interactions originating from electromagnetic fluctuations play a prominent role for many materials and devices. The Casimir force, first predicted by Casimir \cite{Casimir:1948:Main,*Casimir:1948:TIrLdWf}, is of importance especially for nanostructured or layered systems. One particular example is graphite composed of atomically thin graphene layers. The experimental realization of a single graphene \cite{Geim:2007:Trog} has attracted much attention due to its unusual properties and possible practical applications \cite{Katsnelson:2012:GCiTD}. Contemporary experimental techniques enable the synthesis of not only a single graphene, but also a stack of a finite number of graphenes. Since graphene interaction is of fundamental relevance, such progress in the laboratory motivates theoretical studies of how the electromagnetic coupling evolves as a function of the number of stacked layers.

The graphene/graphene and graphene/substrate Casimir interactions have been studied by several authors recently.  Earlier reports have utilized the hydrodynamic model for thin conducting shells \cite{Barton:2004:Cesps,*Barton:2005:CefpsIE,*Bordag:2006:TCeftpsatrotsp,*Bordag:2007:Ioacwatps,*Bordag:2008:Ovesps}, however these results were found to be not suitable  since the hydrodynamic model does not take into account the Dirac-like nature of the carriers  \cite{Klimchitskaya:2015:CohmogwreomtCi}. Other studies have described the graphene response properties via the longitudinal polarization function, optical conductivity, and the polarization operator in a $2+1$ model \cite{Sernelius:2012:Riigs,Drosdoff:2010:Cfgs,Bordag:2009:CibapcagdbtDm,*Fialkovsky:2011:FCefg,*Bordag:2012:TCeitiogwdam,*Bordag:2015:Qftdftrog}. These approaches have solidified our understanding that the Dirac-like carriers coupled with other factors, such as temperature and doping, result in highly unusual behavior of the Casimir force in terms of distance dependence and other asymptotic dependences  \cite{Klimchitskaya:2014:TotCifgsutptacwe,*Klimchitskaya:2014:OoteitCifgs,*Klimchitskaya:2014:TafdtCiigDcfvpt}.

The optical conductivity of graphene $\sigma$  is of particular interest for the Casimir interaction.  It has been shown  \cite{Gusynin:2007:Mcig,*Falkovsky:2007:Sdogc,*Nair:2008:FSCDVToG} that  $\sigma$ takes a constant universal value $(\sigma_{gr} = e^2/4\hbar)$ over a relatively large frequency range, $\hbar\omega \leq 3$ $eV$. This frequency range is commensurate with distance separations in the $\mu m$ and sub-$\mu m$ scale, meaning that $\sigma_{gr}$ is suitable to determine the asymptotic behavior of the graphene Casimir interaction \cite{Drosdoff:2010:Cfgs}. Considering other configurations with an arbitrary value for the constant conductivity has also been of interest, including calculations for the Casimir energy for planar and spherical symmetries  \cite{Khusnutdinov:2014:Cefswcc}. The energy dependence on the separation between two planes  is  $\mathcal{E} \sim 1/d^{3}$, while for a sphere with radius $R$  it is $\mathcal{E} \sim 1/R$. One arrives at these results by simple dimensional considerations. In both cases, taking the conductivity $\sigma$ to be the graphene universal value, $\sigma_{gr}$, leads to a Casimir interaction that does not depend on the Planck constant and the velocity of light in agreement with earlier results  for graphene planes \cite{Drosdoff:2010:Cfgs}. 

The purpose of this paper is to investigate the vdW/Casimir interaction in a stack of $\cN$ infinitely thin parallel layers, as each layer is described by its conductivity $\sigma$. Different models are taken for  $\sigma$ including $\sigma=const$, which can be appropriate for graphene sheet and $\sigma=\infty$, which is suitable for perfect metals. Results for a Drude-Lorentz (DL) model are also derived with the application for graphite parametrization \cite{Djurisic:1999:Opog}. 

We utilize the mode summation method with the zeta-regularization approach for the calculations of the interaction energy and force. The main ingredients of this theory with relevant applications are summarized in \cite{Bordag:2009:ACE}. The energy calculated via the mode summation is a regularized quantity in the following form
\begin{eqnarray}
\mathcal{E}(s) &=& - \hbar c \Lambda^{2s} \frac{\cos\pi s}{2\pi} \iint \frac{d^2k_\perp}{(2\pi)^2} \int_0^\infty d\lambda \lambda^{1-2s}\ann
&\times& \frac{\partial}{ \partial \lambda} \ln \Psi(i\lambda c), \label{eq:EPlanar}
\end{eqnarray}
where $\Psi$ defines the energy spectrum from the relation $\Psi(\omega) = 0$. The parameter $\Lambda$ with a wavenumber dimension is introduced to preserve the energy dimension of $\mathcal{E}(s)$. Note that $\mathcal{E}(s)$ must be renormalized by subtracting $\mathcal{E}_\infty$ corresponding to the  energy for infinitely separated objects ($d\to\infty$) 
\begin{equation}
\mathcal{E}_\infty = \frac{\hbar c}{2\pi} \iint \frac{d^2k_\perp}{(2\pi)^2} \int_0^\infty d\lambda \ln \frac{\Psi(i\lambda c)}{\Psi(i\lambda c)|_{d\to\infty}}. \label{eq:EPlanarMain}
\end{equation} 

The paper is organized as follows. In Sec. \ref{Sec:PlSym} we derive the general expression for Casimir energy for a stack of $\cN$ planes with constant conductivity and analyze them in  detail.  Section \ref{Sec:Force} is devoted to the derivation of the Casimir force acting on a plane in the stack with constant conductivity. In Sec. \ref{Sec:DL} we consider the DL model with parametrization for graphite and analyze expressions obtained for the energy. We show that the binding energy per unit atom is close to the one obtained experimentally. Section \ref{Sec:Concl} contains our conclusions and discussion. In Appendix \ref{Sec:AppA} we make derivation of the expression for the energy of stack of $\cN$ planes. The calculation of the force is made in Appendix \ref{Sec:AppB}. 

Throughout the paper we use units in which $\hbar = c =1$. Where necessary the constants are restored for better understanding of the final results.

\section{Casimir energy in a stack of planes}\label{Sec:PlSym}
\begin{figure}[ht]
	\centerline{\includegraphics[width=7cm]{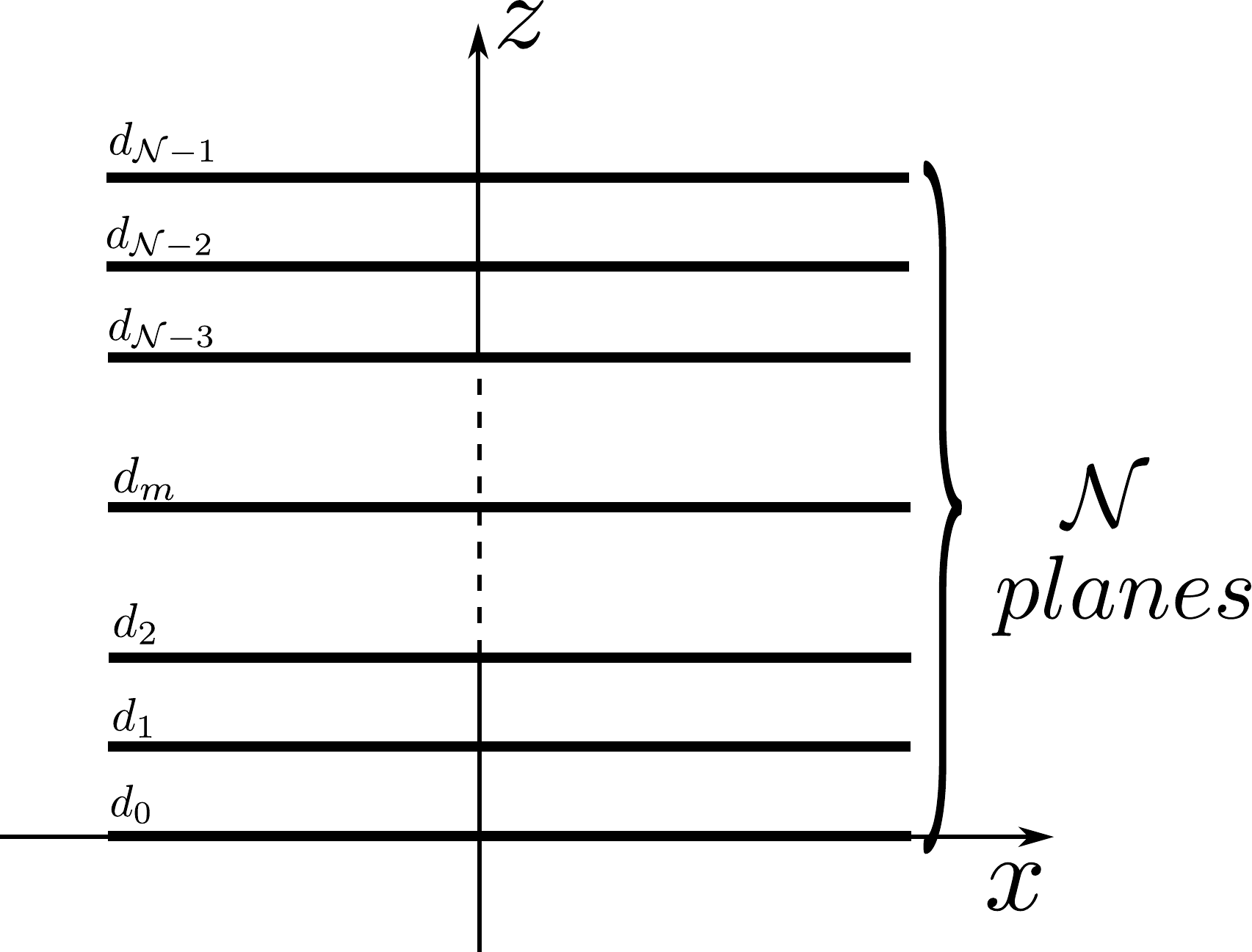}} 
	\caption{A schematic representation of $\cN$ parallel infinitely thin layers separated in the $z$-direction and extending in the $x-y$ plane. The planes are considered to be identical and equally spaced according to $d_m = md$.  }\label{fig:1}
\end{figure}

The system under consideration consists of $\cN$  equally spaced parallel and infinitely thin layers, as shown in Fig. \ref{fig:1}.  

We first calculate the Casimir energy for this system by assuming that all planes have a constant conductivity. For this purpose, the appropriate boundary conditions for the electric and magnetic fields are imposed, which leads to a separation of the transverse electric (TE) and transverse magnetic (TM) modes \cite{Khusnutdinov:2014:Cefswcc}. As a result, the interaction energy $\mathcal{E}^{(\cN)}$ stored in the stack can be represented as
\begin{equation}
\mathcal{E}^{(\cN)} = \frac{Q^{(\cN)}(\eta)}{d^3} = \frac{Q^{(\cN)}_\TM(\eta) + Q^{(\cN)}_\TE(\eta)}{d^3},\label{eq:TotE}
\end{equation}
\begin{eqnarray}
Q^{(\cN)}_\TM(\eta) &=&  \frac{1}{32\pi^2} \int_0^\infty y^2dy\int_0^1 dx \ln F_\cN\left(\frac{\eta}{x}, \frac{y}{2}\right),\ann
Q^{(\cN)}_\TE(\eta) &=&  \frac{1}{32\pi^2} \int_0^\infty y^2dy\int_0^1 dx \ln F_\cN\left(\eta x, \frac{y}{2}\right), \label{eq:TotEQ}
\end{eqnarray}
where $\eta = 2\pi\sigma/c$ is a dimensionless parameter and  the integrand function is 
\begin{gather}
F_\cN (t,z) = - \frac{e^{-z (\cN-1)} }{f^{\cN-2}(1+t)^\cN} \ann 
\times \left( e^{-z} \frac{1-f^{2(\cN-1)}}{1-f^2} - \frac{1+t}{f} \frac{1-f^{2\cN}}{1-f^2}\right), \label{eq:FN}
\end{gather}
where $f = f(t,z) = \sqrt{(\cosh z + t \sinh z)^2 -1} + \cosh z + t \sinh z$. Details of the derivation of the interaction energy can be found in Appendix \ref{Sec:AppA}. It is noted that the cases of $\cN = 3,4$, consistent with Eq. (\ref{eq:TotE}), have been examined in 
Ref. \cite{Khusnutdinov:2015:Cefacopcs}. 

To understand the main factors affecting the Casimir energy, we define the average energy per plane for the stack of planes
of $\cN$ planes  
\begin{equation}
\overline{\mathcal{E}}^{(\cN)} = \frac{\mathcal{E}^{(\cN)}}{\cN} = \frac{\overline{Q}^{(\cN)}(\eta)}{d^3} = \frac{\overline{Q}^{(\cN)}_\TM(\eta) + \overline{Q}^{(\cN)}_\TE(\eta)}{d^3},\label{eq:TotEAve}
\end{equation}
where $\overline{Q}^{(\cN)}(\eta)=Q^{(\cN)}(\eta)/\cN$.

We further consider the energy $\overline{\mathcal{E}}^{(\infty)}$ for infinite number of planes, which can be expressed by taking the $\cN\to\infty$ limit in Eqs. (\ref{eq:TotE}) -- (\ref{eq:TotEAve}). As a result, the function $F_\cN (t,z)$ is replaced by $\Phi(t,z) = \lim_{\cN\to\infty}\sqrt[\cN]{F_{\cN}(t,z)} = e^{-z}f(t,z)/(1+t)$  giving $\overline{Q}_{\TE, \TM}^{(\infty)}$.

\subsection{Calculation of the Casimir energy}

It is instructive to study the case of infinite number of perfectly conducting layers firts. We see that for $\eta \to \infty$, one finds $F_{\cN} \to (1-e^{-2z})^{\cN-1} $ from Eq. (\ref{eq:FN}). Therefore, the energy stored in the stack  is 
\begin{equation}
\mathcal{E}^{(\cN)} = (\cN-1)\mathcal{E}_{Cas},
\end{equation}
where $\mathcal{E}_{Cas} =  -\frac{\pi^2}{720 d^3}$ is the energy between two perfect metallic substrates separated by a distance $d$. Using Eq. (\ref{eq:TotEAve}) for the average energy, it is easy to see that for infinite number of planes, we recover the standard result for the Casimir energy between perfect metals:
\begin{equation}
\lim_{\cN\to\infty} \overline{\mathcal{E}}^{(\cN)}  = \mathcal{E}_{Cas}.
\end{equation}

In addition to $\eta \to \infty$, we also calculate the energy for a small constant conductivity in the limit of  $\eta \to 0$. In this case, the main contribution comes from the TM modes, and the energy per unit plane is of the form
\begin{equation}
\overline{\mathcal{E}}^{(\cN)}  = \mathcal{E}_{Cas} \sigma b_\cN, 
\end{equation}
where 
\begin{equation}
b_{\cN} = -\frac{45}{\cN \pi^3} \int_0^\infty y^2dy\int_0^\infty dx \ln F_\cN\left(\frac{1}{x}, \frac{y}{2}\right). 
\end{equation}

These expressions give the means to evaluate the Casimir energy for different number of layers $\cN$. For example, for two layers ($\cN=2$) and infinite number of layers ($\cN=\infty$), in the $\eta \to \infty$ we find the analytic relation  
\begin{equation}
b_2= 1.4864 ; b_\infty = \pi. 
\end{equation}
Numerical calculations for intermediate cases are provided in Table \ref{tab:1}. It is clear that the $b_\cN$ coefficient plays the role of a scaling constant to the energy. Therefore, the interaction has a universal behavior, in which the energy per unit plane is simply proportional to $\mathcal{E}_{Cas}$ and the small conductivity.
\begin{table}[ht]
	\centerline{\begin{tabular}{|c|c|c|c|}
			\hline 
			$\cN$ & $b_\cN$& $\cN$ & $b_\cN$ \\ 
			\hline 
			2 &  1.48641 & 5 &  2.46766\\ 
			\hline 
			3 &  2.02535 & 6&  2.57937\\  
			\hline 
			4 &  2.30086 & $\infty$ & $\pi$\\ 
			\hline 
		\end{tabular} }\caption{The constants $b_\cN  = \overline{\mathcal{E}}^{(\cN)}/\mathcal{E}_{Cas} \sigma$ in the case of small conductivity $\sigma\to 0$.}\label{tab:1}
	\end{table}

\begin{figure}[ht]
	\centerline{\includegraphics[width=8cm]{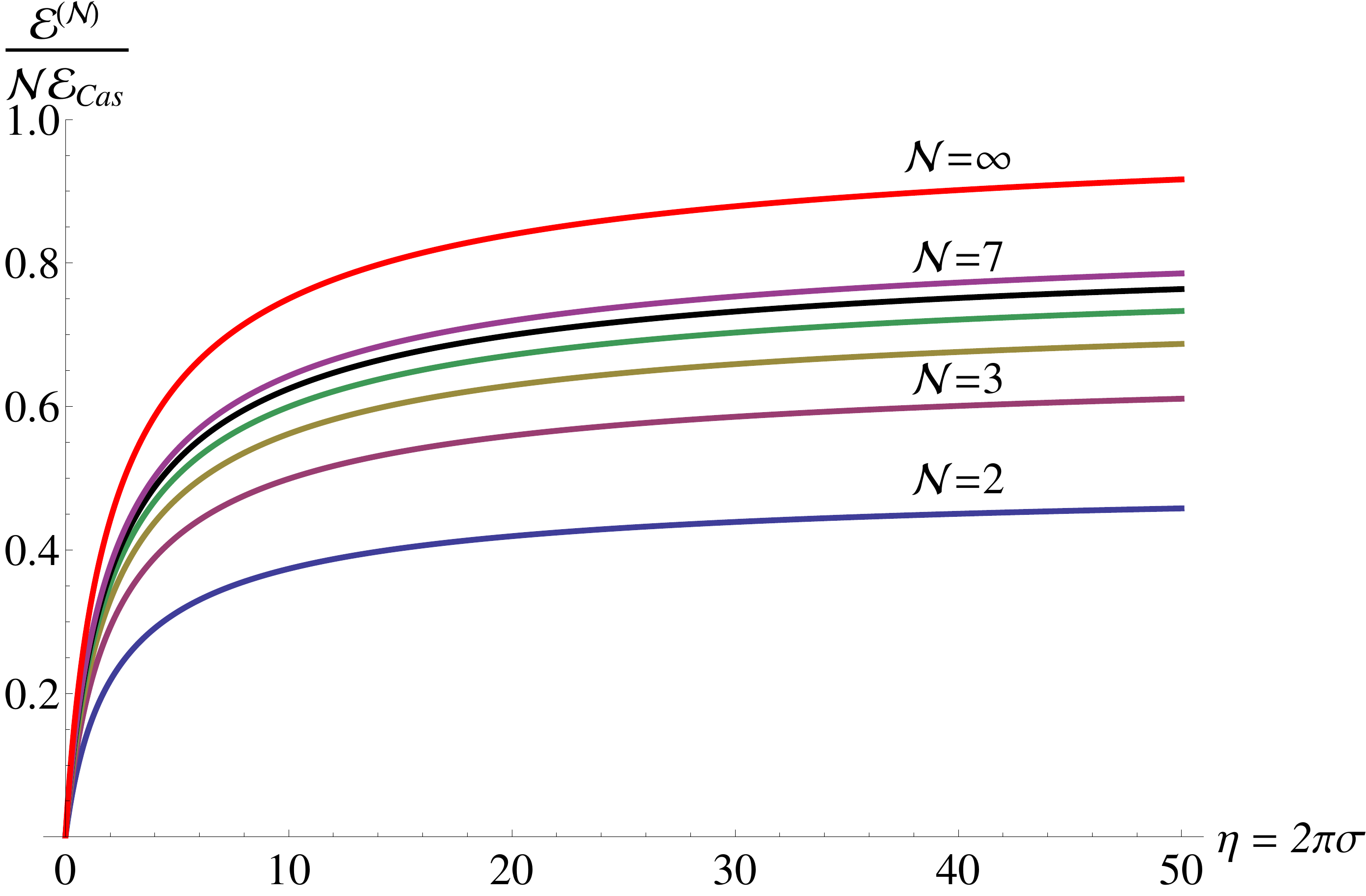}} 
	\caption{The normalized average energy of $\cN$ planes ${\mathcal{E}^{(\cN)}}/{\cN \mathcal{E}_{Cas}}$ as a function of the dimensionless conductivity $\eta = 2\pi  \sigma/c$.}\label{fig:2}
\end{figure}

Numerical simulations help us understand how the energy behaves for intermediate values of $\eta$. In Fig. \ref{fig:2} results are shown for the energy per unit plane normalized by $\mathcal{E}_{Cas}$ for different number of the planes. 
We observe that increasing $\cN$ correlates with larger energy. For infinite number of planes the energy per unit plane approaches the Casimir energy for two ideal metals.  

\section{The force}\label{Sec:Force}
To gain a better perspective of the Casimir interaction, the force on a particular plane in the stack $\cN$ is also considered. This is a convenient way to explicitly understand how the interaction changes depending on the location of the plane. The Casimir force acting on plane $m$ in the stack $\cN$ (Fig.\ref{fig:1}) can be written as (see Appendix \ref{Sec:AppB}) 
\begin{equation}
\mathcal{F}^{(m,\cN)} = \frac{U^{(m,\cN)}(\eta)}{d^4} = \frac{U^{(m,\cN)}_\TM(\eta) + U^{(m,\cN)}_\TE(\eta)}{d^4},\label{eq:Force}
\end{equation}
where the TE and TM contributions  read
\begin{eqnarray*}
U^{(m,\cN)}_\TM(\eta) &=&  \frac{1}{32\pi^2} \int_0^\infty y^2dy\int_0^1 dx  G_{m,\cN}\left(\frac{\eta}{x}, \frac{y}{2}\right),\ann
U^{(m,\cN)}_\TE(\eta) &=&  \frac{1}{32\pi^2} \int_0^\infty y^2dy\int_0^1 dx  G_{m,\cN}\left(\eta x, \frac{y}{2}\right).
\end{eqnarray*}
These are expressed in terms of a single function
\begin{equation}
G_{m,\cN} (t,z) = -\frac{2z t^2f e^{-z}\left(f^{2m}-f^{2(\cN-1-m)}\right)}{e^{-z}f (1-f^{2(\cN-1)}) - (1-f^{2\cN}) (1+t) }, \label{eq:G}
\end{equation}
where $f(t,z)$ was defined in the previous section. Examining the general properties of Eq. (\ref{eq:G}), one finds the following. The function $G$ is odd with respect to its indexes, thus $G_{m,\cN} = - G_{\cN-1-m,\cN}$. Therefore, for two symmetric planes, for which $m\Leftrightarrow \cN -1 -m$, the force has the same magnitude, but opposite sign signaling attraction. Also, the force on the central plane $m = (\cN-1)/2$ in a stack of odd number of plane is zero.

{\it Perfect Metals $\eta \to \infty$}. We first consider the case of  perfect metallic planes. The force on a plane has different characteristic depending on the particular $m$. For the first plane at $m=0$, it is obtained that $\lim_{\eta\to\infty} G_{0,\cN} = -\frac{2z}{e^{2z} - 1}$, which leads to equal TE and TM contributions 
\begin{equation}
\lim_{\eta\to\infty} U^{(0,\cN)}_{\TE,TM} = - \frac{\pi^2}{480} \rightarrow \mathcal{F}^{(0,\cN)} = -\frac{\pi^2}{240 d^4}.
\end{equation} 
Therefore, the force on the very first (or last) plane and the rest of the stack corresponds to the Casimir force between two perfect metals.

For the second plane $m=1$ in the stack, the TM contribution is:
\begin{eqnarray}
\lim_{\eta\to\infty} U^{(1,\cN)}_\TM &=& \frac{1}{32\pi^2 \eta^2} \int_0^\infty y^2dy\int_0^1 dx \frac{y^3 x^2 e^{y}}{(e^y -1)^3}\ann 
&=& \frac{\pi^2 - 6\zeta_R(3)}{192\pi^2 \eta^2}. 
\end{eqnarray}
For the TE modes, however, the domain of integration over $x$ needs to be divided into two parts to avoid the occurring divergences. Thus it is found
\begin{gather}
\lim_{\eta\to\infty} U^{(1,\cN)}_\TE =  \lim_{\eta\to\infty} \frac{1}{32\pi^2} \int_0^\infty y^2dy\int_0^\infty dx  G_{1,\cN}\left(\eta x, \frac{y}{2}\right)\ann 
- \lim_{\eta\to\infty} \frac{1}{32\pi^2} \int_0^\infty y^2dy\int_1^\infty dx  G_{1,\cN}\left(\eta x, \frac{y}{2}\right). \label{eq:U1TE}
\end{gather}
Further examination shows that the first term in the above equation falls down as $\eta^{-1}$ and second term -- as $\eta^{-2}$. Comparing with $\lim_{\eta\to\infty} U^{(1,\cN)}_\TM \sim \eta^{-2}$, it is concluded that the dominant term is the TE term of the order  $\eta^{-1}$. Nevertheless, the force on the second plane will be asymptotically zero: $\mathcal{F}_{\eta\to\infty}^{(1,\cN)}=0$.  

For any other plane $\cN/2 \geq m\geq 2$, the TM contribution is evaluated by changing variables $\ y= \ln (1+ s x/\eta)$ in Eq. (\ref{eq:Force}), after which the limit $\eta \to \infty$ is executed. One obtains 
\begin{equation}
\lim_{\eta\to\infty} U^{m,\cN}_\TM = -\frac{1}{128\pi^2\eta^3}\int_0^\infty ds\frac{s^3 w (w^{2m} - w^{2(\cN-1-m)})}{w^{2\cN} -1}, \label{eq:UmTM}
\end{equation}
where $w = 1+s + \sqrt{s(s+2)}$.  The results for the TE modes are the same as for the $m=1$ plane (Eq. (\ref{eq:U1TE})). We see that  $U^{m,\cN}_\TM$ for any plane with $m\geq 2$ has even stronger dependence on the conductivity as compared to the second one. Nevertheless, dominant term in $U^{m,\cN}_\TE$ is  of the order $\eta^{-1}$ (Eq. (\ref{eq:UmTM})) and it ensures that the force on all planes except the first (or last) one is asymptotically zero. This is a curious result, which says that the dominant interaction present in a stack of perfect metallic planes is between the outer one and the rest of the stack with the Casimir force being the typical one for two perfect metallic substrates. 

{\it Small conductivity $\eta \to 0$}. The case for vanishingly small conductivity is considered next. For this situation, the force on the $m$-th plane is determined primarily by the TM mode according to:
\begin{equation}
\mathcal{F}^{(m,\cN)} =  \frac{\sigma}{16\pi^4 d^4} \int_0^\infty\hspace{-1ex} y^2dy\int_0^\infty dx  G_{m,\cN}\left(\frac{1}{x}, \frac{y}{2}\right). \label{eq:Feta0}
\end{equation}
Evidently the force on any plane in the stack will have the same distance dependence and proportionality to $\sigma$. Its  strength is determined by the numerical value of the above integral for a given $m$ and $\cN$. To quantify how the particular location of the plane affects its Casimir force, we define the following parameter 
 \begin{equation}
s_{m,\cN} = \frac{\mathcal{F}^{(m,\cN)}}{\sigma \mathcal{F}_{Cas}},\label{eq:smn} 
\end{equation}
where $\mathcal{F}_{Cas} = \pi^2/240d^4$ is the Casimir force for two perfect metals. For $\cN=\infty$ and large $m$, we  obtain the following asymptotic expression
\begin{equation}
s_{m \gg 1,\infty} \approx \frac{2880 (\ln 8 -2)}{\pi^3 (2m+1)^4}. \label{eq:sminf}
\end{equation}
showing that the Casimir force falls down quickly as the plane moves away from the end of the stack. This $\sim m^{-4}$ behavior can be compared with the numerical calculations for $s_{m \gg 1,\infty}$ using Eq. (\ref{eq:Feta0}). The first $10$ coefficients are shown in Table \ref{tab:2} for the case of the  infinite stack.
\begin{table}[ht]
		\centerline{\begin{tabular}{||c||c|c||c||c|c|}
				\hline 
				$m$ & $s_{m,\infty}$& $\tilde{s}_{m,\infty}$& $m$ & $s_{m,\infty}$& $\tilde{s}_{m,\infty}$ \\ 
				\hline\hline 
				0 &  3.05 &14.51 & 5 & $4.9\cdot 10^{-4}$ & $1.5\cdot 10^{-3}$\\ 
				\hline 
				1 &  $7.6\cdot 10^{-2}$ &  $2.4\cdot 10^{-1}$ & 6&  $2.5\cdot 10^{-4}$ & $7.7\cdot 10^{-4}$\\  
				\hline 
				2 &  $1.1\cdot 10^{-2}$ & $3.3\cdot 10^{-2}$  & 7 & $1.4\cdot 10^{-4}$ &  $4.3\cdot 10^{-4}$ \\ 
				\hline 
				3 &  $2.9\cdot 10^{-3}$ & $9.1\cdot 10^{-3}$  & 8 & $8.7\cdot 10^{-5}$ & $2.5\cdot 10^{-4}$ \\
				\hline 
				4 &  $1.1\cdot 10^{-3}$ & $3.3\cdot 10^{-3}$ & 9 & $5.6\cdot 10^{-5}$ & $1.5\cdot 10^{-4}$\\
				\hline 
			\end{tabular}}\caption{The constants $s_{m,\infty}  = \mathcal{F}^{(m,\infty)}/\mathcal{F}_{Cas} \sigma$ in the case of small constant conductivity $\sigma\to 0$ and $\tilde{s}_{m,\infty}  = \mathcal{F}^{(m,\infty)}(\tilde{\eta}_{DL})/\mathcal{F}_{Cas} \sigma_{gr}$ for the Drude-Lorentz model.}\label{tab:2}
	\end{table}
We see that the asymptotic expression $s_{m,\infty}$ differs  by mere $7\%$ for $m=2$ from the computational result, while for the seventh plane the error is smaller than $1\%$.

\begin{figure}[ht]
	\includegraphics[width=8cm]{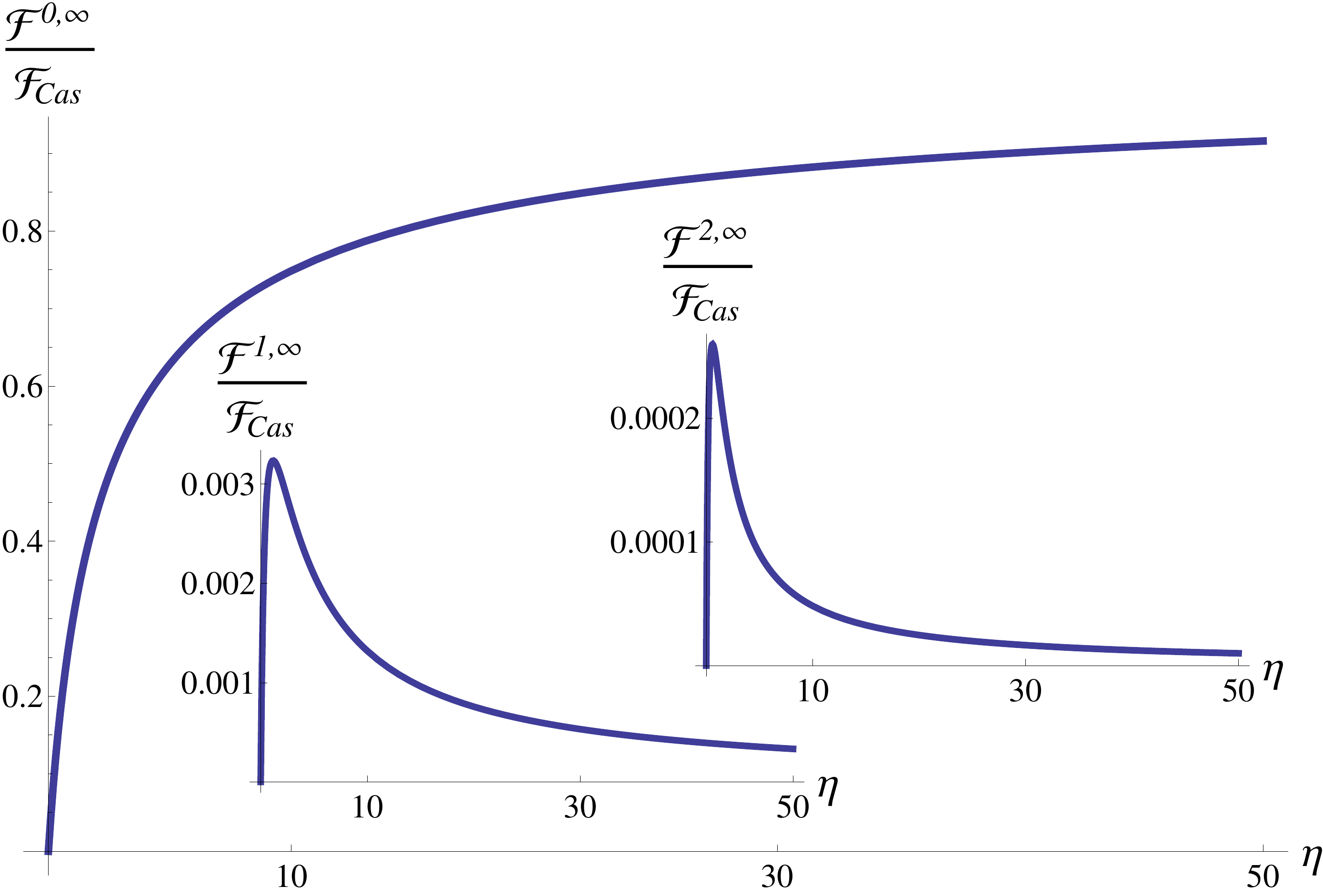}	
	\caption{The Casimir force $\mathcal{F}^{(m,\infty)}$    normalized by $\mathcal{F}_{Cas}=\frac{\pi^2}{240d^4}$  as a function of the dimensionless conductivity $\eta = 2\pi \sigma$.  The inserts show the normalized Casimir force vs. $\eta$ for the $m=1$ and $m=2$ planes. }\label{fig:4}
\end{figure}

{\it Numerical Calculations} For intermediate values of $\eta$, one must examine Eqs. (\ref{eq:Force}) via numerical techniques. In Fig. \ref{fig:4}, we show how the normalized to the perfect metal value force evolves as a function of the conductivity in the case of $\cN=\infty$.  

It is observed that the force starts out as $\sim \eta$ and then asymptotically approaches the value for perfect metals. The force on any other plane ($m=1,2$ only shown), however, exhibits an extreme behavior with a sharp maximum in the small $\eta$ range. Also,  $\mathcal{F}^{(m,\infty)}$ for any $m\geq 1$  is orders of magnitude smaller than $\mathcal{F}^{(0,\infty)}$ showing that the overall Casimir interaction is determined primarily by the  very first plane and the rest of the stack for any value of the conductivity.

\vspace{.5cm}

\section{Drude-Lorentz model of conductivity}\label{Sec:DL}

The Casimir interaction depends strongly on the response properties of the materials. In addition to infinitely thin planes with $\sigma=const$, we also calculate the Casimir interaction within a Drude-Lorentz (DL) model for the conductivity of each plane.  The DL model leads to an interesting asymptotic distance dependence showing once again the richness of this ubiquitous interaction.

As a model, we utilize the optical response of graphite. First principles calculations indicate that the optical conductivity as per graphene sheet is very close to the one for an isolated graphene over a wide range of frequencies \cite{Marinopoulos:2004:isotoaawdrog}. The results for graphene have been mapped to a Drude-Lorentz model consisting of  a Drude term and seven Lorentz oscillators according to \cite{Djurisic:1999:Opog} 
\begin{equation}
\sigma_{DL} (\omega) = \frac{f_0 \omega_p^2}{\gamma_0 - i\omega} + \sum_{j =1}^7\frac{i \omega f_j \omega_p^2}{ \omega^2 - \omega_j^2 + i\omega \gamma_j }.\label{eq:DL}
\end{equation}

In this case, the Casimir interaction is calculated using Eq. (\ref{eq:EPlanarMain}), where the spectral function $\Psi$ is a function of $\sigma_{DL}$, $\Psi (\omega,\sigma (\omega))$. It is realized that 
\begin{equation}
\partial_\omega \ln \Psi = \frac{1}{\Psi} \left(\partial_\omega \Psi + \partial_\sigma \Psi \partial_\omega \sigma\right), 
\end{equation}
has the same simple pole structure as $\sigma_{DL}(\omega)$. The renormalized expression $\partial_\omega \ln \frac{\Psi(\omega,\sigma,d)}{\Psi(\omega,\sigma,\infty)}$, however, has no poles. Therefore, the Casimir energy can be found using Eq. (\ref{eq:EPlanarMain}) together with the formalism in Sec. \ref{Sec:PlSym} by setting $\sigma$ to the DL conductivity. 

The graphene DL conductivity is obtained from  Eq.  (\ref{eq:DL}) for the 3D graphite by multiplying it with $2\pi a/c$ ($a$ is a characteristic distance typically taken as the interplanar distance of graphite). The expression is given on the  imaginary axis $\omega = i \lambda c \ (k=i\lambda)$ as follows:
\begin{equation}
\eta_{DL}(\lambda)  = \frac{\eta_0 \tilde\gamma_0}{\tilde\gamma_0 + \lambda} + \sum_{k = 1}^7\frac{\lambda \eta_k \tilde\gamma_k }{ \lambda^2 + \lambda_k^2 + \lambda \tilde \gamma_k }.
\label{eq:DLcompl}
\end{equation}
Here, $\gamma_k$ is the relaxation time and $\omega_k$ is the characteristic frequency for the $k$-th term. Also, $\tilde{\gamma}_k = \gamma_k/c,\ \lambda_k = \omega_k/c$, and $\eta_k = 2\pi a f_k\omega_p^2/c\gamma_k$.  In Table  \ref{tab:DrDirLor} we show the parameters using the calculated values for graphite \cite{Djurisic:1999:Opog}.
\begin{table}[ht]
	\centerline{ \begin{tabular}{|c|c|c|c||c|c||}\hline\hline
			$\eta_k$ & &$\gamma_k$ & $eV$ & $\omega_k$ & $eV$ \\ \hline\hline
			$\eta_0$ & $0.01712$ &$\gamma_0$ & $6.365$ &- &-  \\ 
			$\eta_1$ & $0.13855$ &$\gamma_1$ & $4.102$ & $\omega_1$ & $0.275$  \\
			$\eta_2$ & $0.05949$ &$\gamma_2$ & $7.328$ & $\omega_2$ & $3.508$   \\
			$\eta_3$ & $0.37991$ &$\gamma_3$ & $1.414$ & $\omega_3$ & $4.451$  \\   
			$\eta_4$ & $0.08462$ &$\gamma_4$ & $0.46$ & $\omega_4$ & $13.591$ \\
			$\eta_5$ & $1.09548$ &$\gamma_5$ & $1.862$ & $\omega_5$ & $14.226$  \\
			$\eta_6$ & $0.30039$ &$\gamma_6$ & $11.922$& $\omega_6$ & $15.55$  \\
			$\eta_7$ & $0.03983$ &$\gamma_7$ & $39.091$& $\omega_7$ & $32.011$\\ \hline   
		\end{tabular}%
		\begin{tabular}{|c|c||c|c||}\hline\hline
			$\tilde\gamma_k$ & $\frac{1}{nm}$ & $\lambda_k$  & $\frac{1}{nm} $\\ \hline\hline
			$\tilde\gamma_0$ & $0.0322$ &- &-  \\ 
			$\tilde\gamma_1$ & $0.0207$ & $\lambda_1$ & $0.0014$  \\
			$\tilde\gamma_2$ & $0.0371$   & $\lambda_2$&$0.0177$   \\
			$\tilde\gamma_3$ & $0.0072$ & $\lambda_3$ & $0.0225$ \\   
			$\tilde\gamma_4$ & $0.0023$ & $\lambda_4$ & $0.0688$  \\
			$\tilde\gamma_5$ & $0.0094$ & $\lambda_5$ & $0.0721$  \\
			$\tilde\gamma_6$ & $0.0604$ & $\lambda_6$ & $0.0788$  \\
			$\tilde\gamma_7$ & $0.1981$ & $\lambda_7$ & $0.1622$  \\ \hline   
		\end{tabular}		
}\caption{Parameters for the Drude-Lorentz model of a graphene sheet in 3D graphite}\label{tab:DrDirLor}
	\end{table}

We note that the optical response in the infrared regime for 3D graphite and an isolated graphene is different. While $\sigma$ for graphite exhibits a Drude-like behavior, the graphene optical conductivity is constant. This difference is attributed to the different electronic structure characteristics for the two systems in this range \cite{Marinopoulos:2004:isotoaawdrog}. To ensure that the $\sigma=const$ feature is captured, the single graphene conductivity $\widetilde{\eta}_{DL}$  is obtained by using a characteristic distance $a=0.224$ $nm$. In addition, we require that $\widetilde{\eta}_{DL}(0)$ coincides with $\eta_{gr}$ at zero frequency as:
\begin{equation}
\widetilde{\eta}_{DL} (\lambda) = \eta_{DL} (\lambda) \frac{\eta_{gr}}{\eta_0}.\label{eq:DLGr}
\end{equation} 

The Casimir interaction energy in a stack of planes characterized with a DL response can now be calculated using Eqs. (\ref{eq:TotE}) --  (\ref{eq:TotEAve}), but with conductivity that is depended upon distance -- $\eta \left(\frac{y x }{2 d}\right)$. Examining the large separation limit $d\to\infty\ (q_j = 2d\tilde \gamma_j \to \infty)$ shows that $\eta_{DL} \to \eta_0\ (\tilde{\eta}_{DL} \to \eta_{gr})$. Therefore, the Casimir energy has the same behavior as the one described in Sec. \ref{Sec:PlSym}.

For small separations $d\to 0\ (q_j\to 0)$, where the planes approach the formation of graphite, we find $\eta \approx  \frac{2d}{xy} \sum_{j=0}^7 \eta_j \tilde \gamma_j \to 0$. The main contribution to the energy comes from the TM mode. Using Eq. (\ref{eq:TotE}), we obtain 
    \begin{gather}
	\mathcal{E}^{(\cN)}_{d\to 0}(\eta_{DL}) = \frac{R_\cN}{d^{5/2}}, \label{eq:d52} \ann
	R_\cN =  \frac{r}{32\pi^2} \int_0^\infty y^{3/2}dy\int_0^\infty \frac{dx}{\sqrt{x}} \ln F_\cN\left(\frac{1}{x}, \frac{y}{2}\right).
	\end{gather}
where $r =  \left(\sum_{j=0}^7 \frac{1}{2}\eta_j\tilde \gamma_j \right)^{1/2}$ is a parameter collecting the characteristic frequencies and oscillator strengths from the DL conductivity. The $d^{-5/2}$ asymptotic distance dependence is a consequence of the frequency dependence of the response properties. We also realize, that all oscillators in the DL model participate in the interaction via the $r$, which reflects the broad band nature of the Casimir phenomenon.

\begin{figure}[ht]
	\includegraphics[width=8truecm]{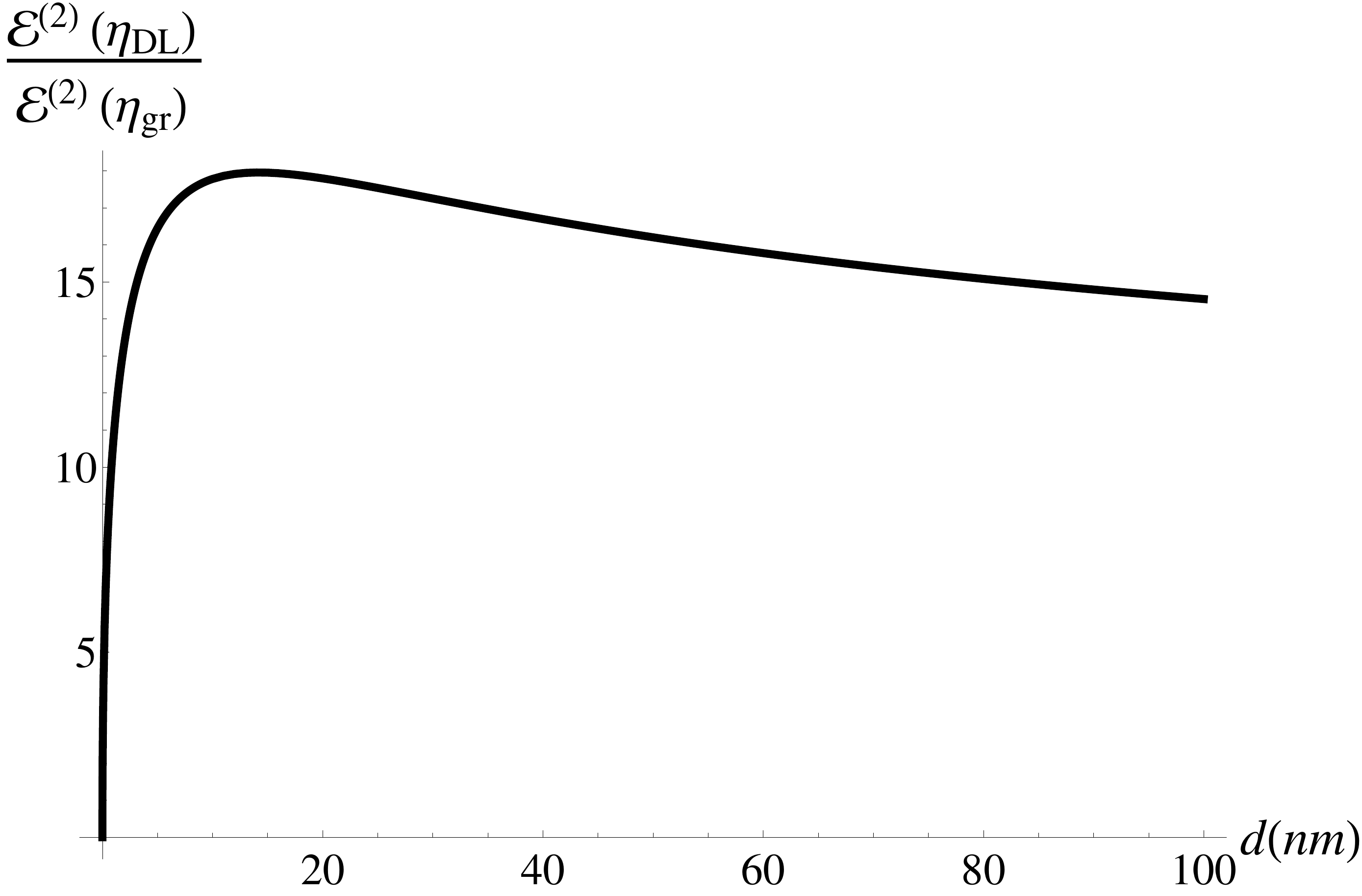}
	\caption{The ratio of the Casimir energy for two planes with dimensionless DL conductivity $\eta_{DL}$ and the Casimir energy for two graphene planes with constant dimensionless conductivity  $\eta_{gr}$.  In the framework of the  DL model, the energy falls down as $d^{-3}$ for large separation and as $d^{-5/2}$ for small separation. The graphene interaction energy has $d^{-3}$ dependence for any distances.}\label{fig:5}
\end{figure}

Therefore, as the planes approach the large and small separation limits, the energy exhibits a transition process in terms of the asymptotic distance dependence. To understand this better, we consider a system from Fig. \ref{fig:1} with $N=2$.  Starting out at $d\to\infty$, where the planes behave as graphene sheets with $\eta = const$, the energy $\mathcal{E} \sim d^{-3}$. As the planes become very close and approach the graphite interplane separation $a=0.334$ $nm$, the conductivity is described by the DL model. Thus the energy exhibits $\mathcal{E} \sim d^{-5/2}$ asymptotic behavior. This transition is shown in Fig. \ref{fig:5} representing plots of the ratio of the Casimir energy for two planes with DL conductivity $\eta_{DL}$ and the Casimir energy for two graphene planes with constant conductivity  $\eta_{gr}$. The maximum of the energy ratio at $d \approx 14$ $nm$ indicates that as $d$ increases, the DL nature of the response becomes less relevant. 

Note that a fractional $d$-dependence of the Casimir interaction was obtained in other systems as well. For example, the energy between two infinitely thin metals has the $d^{-5/2}$ characteristic behavior as found by $ab$ $initio$ simulations \cite{Dobson:2006:ADIABvdWEF,*Dobson:2001:PoDFITaP}. Thin metallic films, described by a Drude model dielectric response, are also characterized by a fractional $d$-dependence \cite{Bostroem:2000:FvdWibtmf}. Based on these results and our investigation, we may conclude that for nanostructured systems described by a DL response, the Casimir interaction is expected to have an unusual, fractional asymptotic distance dependence.

We further consider the average energy per unit plane in an infinite stack of planes using Eq. (\ref{eq:TotE}). Our results show that the separation $a$ plays an important quantitative role for $\mathcal{E}$. While $a=0.334$ $nm$ for the DL model is used for the graphene planes in graphite, $a=0.224$ $nm$ is used to ensure that when graphene planes are separated far apart, the graphene conductivity approaches its constant value $\eta_{gr}$. We obtain that when the plane-plane separation is $d=0.334$ $nm$, $\overline{\mathcal{E}}(\eta_{DL}) = 454.2\frac{erg}{cm^2},\ \overline{\mathcal{E}}(\widetilde{\eta}_{DL}) = 357.1\frac{erg}{cm^2}$ for the two cases, respectively. Describing each plane with a constant conductivity yields $\overline{\mathcal{E}} = 65.2\frac{erg}{cm^2}$ for the same separation. Therefore, the average energy obtained via the DL model is $\approx 5 - 7$ times greater than the one for $\eta_{gr}$.

Estimates for the graphene binding energy in graphite $\mathcal{E}_{ib}=\overline{\mathcal{E}}/a\rho$ can also be given. Here $\rho = 2.83g/cm^3$ is the graphite mass density and $a=0.334$ $nm$ is the graphite interplane separation. We find that   $\mathcal{E}_{ib}(\eta_{DL}) = 74\ meV/atom,\ \mathcal{E}_{ib}(\widetilde{\eta}_{DL}) = 58\ meV/atom$ and for constant graphene conductivity model $\mathcal{E}_{ib}(\eta_{gr}) = 11\ meV/atom$. First principles studies report cohesion energies $24 \div 26\ meV/atom$ \cite{Schabel:1992:Eoibig}, $24\ meV/atom$  \cite{Rydberg:2003:VdWDFLS}. Experimental data shows that the cohesion is $35\pm 10, 15\ meV/atom$ \cite{Benedict:1998:Mdotibeig}, $61\pm 5\ meV/atom$ \cite{Zacharia:2004:Iceogftdoph}. Therefore, our results for graphite sheets described via a constant conductivity underestimate the binding energy, while the DL response ones are fairly in line with the experimental data.

To understand how the interaction evolves as a function of the particular DL plane $m$ in the stack (Fig. \ref{fig:1}), the Casimir force is considered using Eqs. (\ref{eq:Force}), (\ref{eq:G}). The $\tilde{s}_{m,\cN}$ from Eq. (\ref{eq:smn}) with $\sigma = \sigma_{gr}$ not only depends on $m$ but also on the interplanar distance $d$, which is unlike the constant conductivity case.  Furthermore, the force on the $m$-th plane is a function of the combination $(2m+1)\tilde\eta$, which can be used to derive asymptotic expressions in terms of the $m$-th dependence. It is found that for $(2m+1)\tilde{\eta}_{max} \gg 1$ witht $m>3$ and $\tilde\eta_{max} = 0.177$  corresponding to the maximum of the DL conductivity, the following expressions hold:
\begin{eqnarray}
\tilde{s}_{m\gg 1,\infty}^{TM} &\approx& \frac{37800}{\pi^3 \eta_{max}^3\eta_{gr}(2m+1)^{8}} = \left(\frac{8.127}{2m+1}\right)^8,\ann
\tilde{s}_{m\gg 1,\infty}^{TE} &\approx& \frac{1200}{\pi^3 \eta_{max}\eta_{gr}(2m+1)^{6}} =  \left(\frac{5.167}{2m+1}\right)^6. \label{eq:Alpha}
\end{eqnarray}
Therefore, the large $m$ behavior is determined primarily by the TE mode contribution according to Eqs. (\ref{eq:Alpha}). To compare with the results for small $m$, we give the first 10 values of $\tilde{s}_{m,\infty}$ in Table \ref{tab:2}. It is evident that the magnitude of $\mathcal{F}$ decreases rapidly beyond the first few planes in the stack, similar to the case of constant conductivity.


\section{Conclusion} \label{Sec:Concl} 

In summary, we have applied the zero-frequency mode summation method to calculate the Casimir interaction in a system composed of infinitely thin equally spaced layers. The obtained boundary conditions lead to separating the TE and TM mode contributions, which is typical for planar configurations. The response properties are considered within a constant conductivity and Drude-Lorentz models. 

The energy stored in the stack of planes characterized with constant conductivity is found to be  $\overline{\mathcal{E}}^{(\cN)}\sim 1/d^3$ with a magnitude dependent on $\sigma$ and the number of planes. While for perfect conductors, the usual Casimir energy for two infinitely conducting metals is recovered, $\overline{\mathcal{E}}^{(\cN)}\sim \sigma$ for  small conductivity planes. Interestingly, the Casimir force has a very strong dependence on the location of the plane in the stack. The attraction is strongest between the first few planes and the rest of the stack and it decreases by orders of magnitude for larger $m$.

The DL model changes dramatically the asymptotic distance dependence of the interaction especially at layer separation $d< 5nm$. The energy has a fractional dependence according to $\overline{\mathcal{E}}^{(\cN)}\sim 1/d^{5/2}$. The characteristics of all oscillators, quantified in the parameter $r$, enter the energy expressions indicating that all oscillators are important. Also, using the specific values for the graphite DL conductivity, we are able to obtain binding energies consistent with reported experimental data.

This study broadens the application of the zero-point summation method beyond the typical Casimir interactions between two objects of planar symmetry. It shows that this particular approach can be used not only in systems with constant conductivity, but also in systems with frequency dependent response. Our results present further evidence of the complexity of this phenomenon in terms of the nontrivial effects and importance of geometry and response of the interaction objects.

\begin{acknowledgements}
	NK and RK were supported in part by the Russian Foundation for Basic Research Grant No. 13-02-00757-a. LMW acknowledges financial support from the US Department of Eenergy under Contract No. DE-FG02-06ER46297.
\end{acknowledgements}

\appendix
\section{Interaction energy stored in the stack of $\cN$ planes}\label{Sec:AppA}

There are $\cN -1$ interplanar regions for the stack of $\cN$ planes, and the vacuum modes in each region are defined by two constants for a total of $2(\cN-1)$ constants. The spectrum outside of the stack is defined by a single constant in each semi-space. Therefore, there are $2\cN$ constants to be defined. At the same time, we have  $2\cN$ equations for the modes (see Ref.  \cite{Khusnutdinov:2014:Cefswcc}), which can be found from the appropriate boundary conditions on a plane at a position $z=d_j$ for the TM mode  
\begin{eqnarray}
\left[e_z'\right]_{z = d_j} &=& 0,\ann
\left[e_z\right]_{z = d_j} &=& -\frac{4\pi i \sigma }{\omega} e'_z|_{z = d_j},\label{eq:TMmode}
\end{eqnarray}
and the TE mode
\begin{eqnarray}
\left[h_z\right]_{z = d_j} &=& 0,\ann
\left[h_z'\right]_{z = d_j} &=& 4\pi i \sigma \omega h_z|_{z = d_j},\label{eq:TEmode}
\end{eqnarray}
with the electromagnetic fields being 
\begin{equation}
\vec E = \vec e(z) e^{ik_x x + ik_y y - i \omega t},\ \vec H = \vec h(z) e^{ik_x x+ ik_y y - i \omega t},
\end{equation} 
Here $\vec e(z), \vec h(z)$ are the fields amplitudes, which depend on the $z$ coordinate only and  $[f(z)]_a = f(a-0) - f(a+0)$. Each type of mode is considered separately below.

\textbf{TM mode.} The main determinant of the system of equations Eq. (\ref{eq:TMmode}) is expressed in terms of the following matrix 
\begin{gather}
\cA_j =
\begin{pmatrix}
-e^{p_j} & e^{-p_j} \\
(-b-1) e^{p_j} & (b-1) e^{-p_j}
\end{pmatrix},
\cZ_1=
\begin{pmatrix}
-1 & 0 \\
-b-1 & 0 
\end{pmatrix},\ann \ann
\cB_j=
\begin{pmatrix} 
e^{p_{j-1}} & -e^{-p_{j-1}} \\
e^{p_{j-1}} & e^{-p_{j-1}}
\end{pmatrix}, 
\cZ_2=
\begin{pmatrix}
0 & -e^{-p_{\cN-1}} \\
0 & e^{-p_{\cN-1}}
\end{pmatrix},
\end{gather}
where $p_j = d_j \kappa$, $\kappa = \sqrt{k^2_x + k_y^2 + \lambda^2}$, and $\omega = i \lambda $. Also, $b= 2{\eta\kappa}/{\lambda}$ and $\eta=2\pi\sigma$ ($={2\pi\sigma }/{c}$ in dimensional variables).  The determinant has the following $\cN\times \cN$ block matrix:
\begin{equation}
\triangle =
\begin{vmatrix}
\cZ_1 & \cB_1 & 0&\dots& 0&0 \\
0 & \cA_1& \cB_2 &\dots& 0  &0\\
0 & 0 & \cA_2 &\dots& 0 &0\\
\vdots&\vdots&\vdots&\ddots&\vdots&\vdots\\
0 & 0 & 0 &\dots& \cA_{\cN-2} & \cB_{\cN-1} \\
\cZ_2 & 0 & 0 &\dots& 0 & \cA_{\cN-1}
\end{vmatrix}.
\end{equation}

The determinant is then transformed in a diagonal block form   
\begin{equation}
\triangle =
\begin{vmatrix}
\cZ_3 & \cB_1 & 0&\dots& 0&0 \\
0 & \cA_1& \cB_2 &\dots& 0  &0\\
0 & 0 & \cA_2 &\dots& 0 &0\\
\vdots&\vdots&\vdots&\ddots&\vdots&\vdots\\
0 & 0 & 0 &\dots& \cA_{\cN-2} & \cB_{\cN-1} \\
0 & 0 & 0 &\dots& 0 & \cA_{\cN-1}
\end{vmatrix},
\end{equation}
where
\begin{equation}
\cZ_3 = \cZ_1-(-1)^{n}\prod_{j=1}^{n-1}(\cB_j\cA^{-1}_j)\cZ_2.
\end{equation}
This enables writing $\triangle$ in the following way:
\begin{equation}
\triangle =2^{\cN-1}\det\left[ \cZ_3\right], \label{eq:Apdeterminant}
\end{equation}
where we have taken into account that $\det \cA_j = 2$. This factor is omitted in what follows as it does not change the energy spectrum.

Taking into account that the planes are equidistant, the determinant can be cast a slightly different form:
\begin{equation}
\triangle = \det\left[ Z_1-(-1)^{\cN}C^{(\cN-1)}Z_2\right],
\end{equation}
where $p=d\kappa$ and
\begin{equation}
C = B_jA^{-1}_j = 
\begin{pmatrix}
-\cosh p  - b \sinh p & \sinh p \\
b \cosh p+\sinh p & -\cosh p
\end{pmatrix}. \label{eq:ApC}
\end{equation}
The $\cN-1$ power of the matrix $C$ is executed by a Jordan form representation $C=T J T^{-1}$ with 
\begin{gather*}
T= 
\begin{pmatrix}
\frac{\cosh p + f_- }{\sinh p +b \cosh p} & \frac{\cosh p+f_+}{\sinh p + b \cosh p} \\
1 & 1
\end{pmatrix}, \ 
J= 
\begin{pmatrix}
f_- & 0 \\
0 & f_+
\end{pmatrix},
\end{gather*}
where $f_{\pm}=\pm\sqrt{(\cosh p + \frac{b}{2} \sinh p)^2 -1} -(\cosh p + \frac{b}{2}  \sinh p)$.  Therefore, we arrive at
\begin{equation*}
\triangle = \frac{e^{-p(\cN-1)}}{f^{\cN-2}}\left( e^{-p} \frac{1 - f^{2\cN-2}}{1 - f^2} - \frac{1+\frac{b}{2}}{f}\frac{1 - f^{2\cN}}{1 - f^2}\right),
\end{equation*}
where $f = |f_-| = \sqrt{(\cosh p + \frac{b}{2} \sinh p)^2 -1} + \cosh p + \frac{b}{2}  \sinh p$.

To renormalize the spectrum, the limit $d\to \infty\ (p\to\infty)$ is calculated resulting in
\begin{equation}
\triangle_0=\lim_{p\to\infty}\triangle = -\left(1+\frac{b}{2}\right)^\cN.
\end{equation}
The renormalized determinant reads
\begin{gather}
F_\cN\left(\frac{b}{2},p\right) =  \frac{\triangle}{\triangle_0} = -\frac{e^{-p(\cN-1)}}{(1+b/2)^\cN f^{\cN-2}} \ann 
\times \left( e^{-p} \frac{1 - f^{2(\cN-1)}}{1 - f^2} - \frac{1+\frac{b}{2}}{f}\frac{1 - f^{2\cN}}{1 - f^2}\right). 
\end{gather}

To obtain the energy (see Eq. (\ref{eq:EPlanarMain})) the following integration is needed
\begin{equation}
\iint_{-\infty}^\infty \frac{dk_x dk_y}{4\pi^2} \int_0^\infty d\lambda \ln F_\cN. 
\end{equation}
Changing the integrand variables to spherical coordinates $k_x = \kappa \sin\theta \cos\varphi, k_y = \kappa \sin\theta \sin\varphi, \lambda = \kappa \cos\theta\ (\kappa = \sqrt{k_x^2 + k_y^2 + \lambda^2})$, $y = 2p = 2d\kappa, \lambda = \kappa x, x=\cos\theta$  (note that $b/2 = \eta/x$) the TM contribution to the energy becomes
\begin{equation}
\mathcal{E}^{(\cN)}_\TM =  \frac{Q^{(\cN)}_\TM(\eta)}{d^3}, \label{eq:ENtm}
\end{equation}
where  
\begin{equation}
Q^{(\cN)}_\TM(\eta) =  \frac{1}{32\pi^2} \int_0^\infty y^2dy\int_0^1 dx \ln F_\cN\left(\frac{\eta}{x}, \frac{y}{2}\right).
\end{equation}
and the dimensionless function $ F_\cN(t,z)$ is given in Eq. (\ref{eq:FN}).

\textbf{TE mode.} The spectrum calculation for the TE mode follows the same path with the substitution $b = 2\eta \kappa/\lambda \to  2\eta \lambda/\kappa$.   Thus the TE energy is found as 
\begin{equation}
\mathcal{E}^{(\cN)}_\TE =  \frac{Q^{(\cN)}_\TE(\eta)}{d^3},\label{eq:ENte}
\end{equation}
where  
\begin{equation}
Q^{(\cN)}_\TE(\eta) =  \frac{1}{32\pi^2} \int_0^\infty y^2dy\int_0^1 dx \ln F_\cN\left(\eta x, \frac{y}{2}\right).
\end{equation}

\section{Force on the $m$-plane}\label{Sec:AppB}

The Casimir force upon a particular plane $m$ can be calculated using Eqs. (\ref{eq:Apdeterminant}), (\ref{eq:ENtm}) and (\ref{eq:ENte}) by taking an appropriate derivative with respect to $d_m$. For this purpose, we consider the following ratio $\lim_{\epsilon\to 0}(\ln\triangle((m+\epsilon)d) - \ln\triangle(md))/\epsilon$. The matrix $\cZ_3 $ can be regrouped by realizing that $d_m$ appears in $\cA_m$ and $\cB_{m+1}$ only,
\begin{gather}
\cZ_3 = Z_1 - (-1)^{\cN} \left\{\prod_{j=1}^{m}(B_j A^{-1}_j)\right\}\ann 
\times \left\{(B_m \cA^{-1}_m)(\cB_{m+1} A_{m+1}^{-1})\right\}\ann 
\times \left\{\prod_{j=m+1}^{\cN-1}(B_j A^{-1}_j)\right\}Z_2.
\end{gather}
Taking $d_m=(m+\epsilon)d$, one can expand in first order of the small parameter $\epsilon$
\begin{equation}
(B_m \cA^{-1}_m)(\cB_{m+1} A_{m+1}^{-1})  = 
\begin{pmatrix}
1&0\\
0&1
\end{pmatrix} + \varepsilon bp 
\begin{pmatrix}
1&0\\
b&-1
\end{pmatrix}. 
\end{equation}

Thus we can write the following relation 
\begin{gather}
\triangle((m+\epsilon)d) = \det (M + \varepsilon bp P) = \triangle \det(1 + \varepsilon bp M^{-1} P)\ann
=  \triangle (1 + \epsilon b\kappa \Tr (M^{-1} P)).
\end{gather}
where 
\begin{eqnarray*}
	M &=& Z_1-(-1)^{\cN}\prod_{j=1}^{\cN-1}(B_j A^{-1}_j)Z_2\ann
	&=&  Z_1-(-1)^{\cN}C^{\cN-1}Z_2,\\
	P &=& - (-1)^{\cN} C^m 
	\begin{pmatrix}
		1&0\\
		b&-1
		\end{pmatrix}
		C^{\cN-1-m}Z_2,
\end{eqnarray*}
and the matrix $C$ is calulated in (\ref{eq:ApC}). Thus we find
		\begin{gather}
		\lim_{\epsilon\to 0}\frac{\ln\triangle_{m+\epsilon} - \ln\triangle}{\epsilon} = b\kappa  \Tr (PM^{-1})\ann
		= \frac{1}{2}\frac{\kappa b^2 f e^{-p}\left(f^{2m}-f^{2(\cN-1-m)}\right)}{e^{-p}f (1-f^{2(\cN-1)})-(1-f^{2\cN})  \left(1+\frac{b}{2}\right) }.
		\end{gather}
In the limit of $d\to \infty\ (p\to\infty)$ the above expression tends to zero meaning that there is no need for renormalization. Changing the integrand variables to spherical coordinates as previously done in Appendix \ref{Sec:AppA}, Eqs. (\ref{eq:Force}), (\ref{eq:G}) are obtained.

\input{npl.bbl.tex}
\end{document}

%% file: npl.bbl.tex
%

%% file: npl.bbl
\begin{thebibliography}{34}%
\makeatletter
\providecommand \@ifxundefined [1]{%
 \@ifx{#1\undefined}
}%
\providecommand \@ifnum [1]{%
 \ifnum #1\expandafter \@firstoftwo
 \else \expandafter \@secondoftwo
 \fi
}%
\providecommand \@ifx [1]{%
 \ifx #1\expandafter \@firstoftwo
 \else \expandafter \@secondoftwo
 \fi
}%
\providecommand \natexlab [1]{#1}%
\providecommand \enquote  [1]{``#1''}%
\providecommand \bibnamefont  [1]{#1}%
\providecommand \bibfnamefont [1]{#1}%
\providecommand \citenamefont [1]{#1}%
\providecommand \href@noop [0]{\@secondoftwo}%
\providecommand \href [0]{\begingroup \@sanitize@url \@href}%
\providecommand \@href[1]{\@@startlink{#1}\@@href}%
\providecommand \@@href[1]{\endgroup#1\@@endlink}%
\providecommand \@sanitize@url [0]{\catcode `\\12\catcode `\$12\catcode
  `\&12\catcode `\#12\catcode `\^12\catcode `\_12\catcode `\%12\relax}%
\providecommand \@@startlink[1]{}%
\providecommand \@@endlink[0]{}%
\providecommand \url  [0]{\begingroup\@sanitize@url \@url }%
\providecommand \@url [1]{\endgroup\@href {#1}{\urlprefix }}%
\providecommand \urlprefix  [0]{URL }%
\providecommand \Eprint [0]{\href }%
\providecommand \doibase [0]{http://dx.doi.org/}%
\providecommand \selectlanguage [0]{\@gobble}%
\providecommand \bibinfo  [0]{\@secondoftwo}%
\providecommand \bibfield  [0]{\@secondoftwo}%
\providecommand \translation [1]{[#1]}%
\providecommand \BibitemOpen [0]{}%
\providecommand \bibitemStop [0]{}%
\providecommand \bibitemNoStop [0]{.\EOS\space}%
\providecommand \EOS [0]{\spacefactor3000\relax}%
\providecommand \BibitemShut  [1]{\csname bibitem#1\endcsname}%
\let\auto@bib@innerbib\@empty
\bibitem [{\citenamefont {Casimir}(1948)}]{Casimir:1948:Main}%
  \BibitemOpen
  \bibfield  {author} {\bibinfo {author} {\bibfnamefont {H.~B.~G.}\
  \bibnamefont {Casimir}},\ }\href@noop {} {\bibfield  {journal} {\bibinfo
  {journal} {Proc. K. Ned. Akad. Wet.}\ }\textbf {\bibinfo {volume} {51}},\
  \bibinfo {pages} {793} (\bibinfo {year} {1948})}\BibitemShut {NoStop}%
\bibitem [{\citenamefont {Casimir}\ and\ \citenamefont
  {Polder}(1948)}]{Casimir:1948:TIrLdWf}%
  \BibitemOpen
  \bibfield  {author} {\bibinfo {author} {\bibfnamefont {H.~B.~G.}\
  \bibnamefont {Casimir}}\ and\ \bibinfo {author} {\bibfnamefont
  {D.}~\bibnamefont {Polder}},\ }\href {\doibase 10.1103/PhysRev.73.360}
  {\bibfield  {journal} {\bibinfo  {journal} {Phys. Rev.}\ }\textbf {\bibinfo
  {volume} {73}},\ \bibinfo {pages} {360} (\bibinfo {year} {1948})}\BibitemShut
  {NoStop}%
\bibitem [{\citenamefont {Geim}\ and\ \citenamefont
  {Novoselov}(2007)}]{Geim:2007:Trog}%
  \BibitemOpen
  \bibfield  {author} {\bibinfo {author} {\bibfnamefont {A.~K.}\ \bibnamefont
  {Geim}}\ and\ \bibinfo {author} {\bibfnamefont {K.~S.}\ \bibnamefont
  {Novoselov}},\ }\href {\doibase 10.1038/nmat1849} {\bibfield  {journal}
  {\bibinfo  {journal} {Nat. Mater.}\ }\textbf {\bibinfo {volume} {6}},\
  \bibinfo {pages} {183} (\bibinfo {year} {2007})}\BibitemShut {NoStop}%
\bibitem [{\citenamefont {Katsnelson}(2012)}]{Katsnelson:2012:GCiTD}%
  \BibitemOpen
  \bibfield  {author} {\bibinfo {author} {\bibfnamefont {M.}~\bibnamefont
  {Katsnelson}},\ }\href@noop {} {\emph {\bibinfo {title} {Graphene: Carbon in
  Two Dimensions}}}\ (\bibinfo  {publisher} {Cambridge University Press},\
  \bibinfo {year} {2012})\ p.\ \bibinfo {pages} {351}\BibitemShut {NoStop}%
\bibitem [{\citenamefont {Barton}(2004)}]{Barton:2004:Cesps}%
  \BibitemOpen
  \bibfield  {author} {\bibinfo {author} {\bibfnamefont {G.}~\bibnamefont
  {Barton}},\ }\href {\doibase 10.1088/0305-4470/37/3/032} {\bibfield
  {journal} {\bibinfo  {journal} {J. Phys. A: Math. Gen.}\ }\textbf {\bibinfo
  {volume} {37}},\ \bibinfo {pages} {1011} (\bibinfo {year}
  {2004})}\BibitemShut {NoStop}%
\bibitem [{\citenamefont {Barton}(2005)}]{Barton:2005:CefpsIE}%
  \BibitemOpen
  \bibfield  {author} {\bibinfo {author} {\bibfnamefont {G.}~\bibnamefont
  {Barton}},\ }\href {\doibase 10.1088/0305-4470/38/13/013} {\bibfield
  {journal} {\bibinfo  {journal} {J. Phys. A}\ }\textbf {\bibinfo {volume}
  {38}},\ \bibinfo {pages} {2997} (\bibinfo {year} {2005})}\BibitemShut
  {NoStop}%
\bibitem [{\citenamefont {Bordag}(2006)}]{Bordag:2006:TCeftpsatrotsp}%
  \BibitemOpen
  \bibfield  {author} {\bibinfo {author} {\bibfnamefont {M.}~\bibnamefont
  {Bordag}},\ }\href {\doibase 10.1088/0305-4470/39/21/S08} {\bibfield
  {journal} {\bibinfo  {journal} {J. Phys. A}\ }\textbf {\bibinfo {volume}
  {39}},\ \bibinfo {pages} {6173} (\bibinfo {year} {2006})}\BibitemShut
  {NoStop}%
\bibitem [{\citenamefont {Bordag}(2007)}]{Bordag:2007:Ioacwatps}%
  \BibitemOpen
  \bibfield  {author} {\bibinfo {author} {\bibfnamefont {M.}~\bibnamefont
  {Bordag}},\ }\href {\doibase 10.1103/PhysRevD.76.065011} {\bibfield
  {journal} {\bibinfo  {journal} {Phys. Rev. D}\ }\textbf {\bibinfo {volume}
  {76}},\ \bibinfo {pages} {065011} (\bibinfo {year} {2007})}\BibitemShut
  {NoStop}%
\bibitem [{\citenamefont {Bordag}\ and\ \citenamefont
  {Khusnutdinov}(2008)}]{Bordag:2008:Ovesps}%
  \BibitemOpen
  \bibfield  {author} {\bibinfo {author} {\bibfnamefont {M.}~\bibnamefont
  {Bordag}}\ and\ \bibinfo {author} {\bibfnamefont {N.~R.}\ \bibnamefont
  {Khusnutdinov}},\ }\href {\doibase 10.1103/PhysRevD.77.085026} {\bibfield
  {journal} {\bibinfo  {journal} {Phys. Rev. D}\ }\textbf {\bibinfo {volume}
  {77}},\ \bibinfo {pages} {085026} (\bibinfo {year} {2008})}\BibitemShut
  {NoStop}%
\bibitem [{\citenamefont {Klimchitskaya}\ and\ \citenamefont
  {Mostepanenko}(2015)}]{Klimchitskaya:2015:CohmogwreomtCi}%
  \BibitemOpen
  \bibfield  {author} {\bibinfo {author} {\bibfnamefont {G.~L.}\ \bibnamefont
  {Klimchitskaya}}\ and\ \bibinfo {author} {\bibfnamefont {V.~M.}\ \bibnamefont
  {Mostepanenko}},\ }\href {\doibase 10.1103/PhysRevB.91.045412} {\bibfield
  {journal} {\bibinfo  {journal} {Phys. Rev. B}\ }\textbf {\bibinfo {volume}
  {91}},\ \bibinfo {pages} {045412} (\bibinfo {year} {2015})}\BibitemShut
  {NoStop}%
\bibitem [{\citenamefont {Sernelius}(2012)}]{Sernelius:2012:Riigs}%
  \BibitemOpen
  \bibfield  {author} {\bibinfo {author} {\bibfnamefont {B.~E.}\ \bibnamefont
  {Sernelius}},\ }\href {\doibase 10.1103/PhysRevB.85.195427} {\bibfield
  {journal} {\bibinfo  {journal} {Phys. Rev. B}\ }\textbf {\bibinfo {volume}
  {85}},\ \bibinfo {pages} {195427} (\bibinfo {year} {2012})}\BibitemShut
  {NoStop}%
\bibitem [{\citenamefont {Drosdoff}\ and\ \citenamefont
  {Woods}(2010)}]{Drosdoff:2010:Cfgs}%
  \BibitemOpen
  \bibfield  {author} {\bibinfo {author} {\bibfnamefont {D.}~\bibnamefont
  {Drosdoff}}\ and\ \bibinfo {author} {\bibfnamefont {L.~M.}\ \bibnamefont
  {Woods}},\ }\href {\doibase 10.1103/PhysRevB.82.155459} {\bibfield  {journal}
  {\bibinfo  {journal} {Phys. Rev. B}\ }\textbf {\bibinfo {volume} {82}},\
  \bibinfo {pages} {155459} (\bibinfo {year} {2010})}\BibitemShut {NoStop}%
\bibitem [{\citenamefont {Bordag}\ \emph
  {et~al.}(2009{\natexlab{a}})\citenamefont {Bordag}, \citenamefont
  {Fialkovsky}, \citenamefont {Gitman},\ and\ \citenamefont
  {Vassilevich}}]{Bordag:2009:CibapcagdbtDm}%
  \BibitemOpen
  \bibfield  {author} {\bibinfo {author} {\bibfnamefont {M.}~\bibnamefont
  {Bordag}}, \bibinfo {author} {\bibfnamefont {I.~V.}\ \bibnamefont
  {Fialkovsky}}, \bibinfo {author} {\bibfnamefont {D.~M.}\ \bibnamefont
  {Gitman}}, \ and\ \bibinfo {author} {\bibfnamefont {D.~V.}\ \bibnamefont
  {Vassilevich}},\ }\href {\doibase 10.1103/PhysRevB.80.245406} {\bibfield
  {journal} {\bibinfo  {journal} {Phys. Rev. B}\ }\textbf {\bibinfo {volume}
  {80}},\ \bibinfo {pages} {245406} (\bibinfo {year}
  {2009}{\natexlab{a}})}\BibitemShut {NoStop}%
\bibitem [{\citenamefont {Fialkovsky}\ \emph {et~al.}(2011)\citenamefont
  {Fialkovsky}, \citenamefont {Marachevsky},\ and\ \citenamefont
  {Vassilevich}}]{Fialkovsky:2011:FCefg}%
  \BibitemOpen
  \bibfield  {author} {\bibinfo {author} {\bibfnamefont {I.~V.}\ \bibnamefont
  {Fialkovsky}}, \bibinfo {author} {\bibfnamefont {V.~N.}\ \bibnamefont
  {Marachevsky}}, \ and\ \bibinfo {author} {\bibfnamefont {D.~V.}\ \bibnamefont
  {Vassilevich}},\ }\href {\doibase 10.1103/PhysRevB.84.035446} {\bibfield
  {journal} {\bibinfo  {journal} {Phys. Rev. B}\ }\textbf {\bibinfo {volume}
  {84}},\ \bibinfo {pages} {035446} (\bibinfo {year} {2011})}\BibitemShut
  {NoStop}%
\bibitem [{\citenamefont {Bordag}\ \emph {et~al.}(2012)\citenamefont {Bordag},
  \citenamefont {Klimchitskaya},\ and\ \citenamefont
  {Mostepanenko}}]{Bordag:2012:TCeitiogwdam}%
  \BibitemOpen
  \bibfield  {author} {\bibinfo {author} {\bibfnamefont {M.}~\bibnamefont
  {Bordag}}, \bibinfo {author} {\bibfnamefont {G.~L.}\ \bibnamefont
  {Klimchitskaya}}, \ and\ \bibinfo {author} {\bibfnamefont {V.~M.}\
  \bibnamefont {Mostepanenko}},\ }\href {\doibase 10.1103/PhysRevB.86.165429}
  {\bibfield  {journal} {\bibinfo  {journal} {Phys. Rev. B}\ }\textbf {\bibinfo
  {volume} {86}},\ \bibinfo {pages} {165429} (\bibinfo {year}
  {2012})}\BibitemShut {NoStop}%
\bibitem [{\citenamefont {Bordag}\ \emph {et~al.}(2015)\citenamefont {Bordag},
  \citenamefont {Klimchitskaya}, \citenamefont {Mostepanenko},\ and\
  \citenamefont {Petrov}}]{Bordag:2015:Qftdftrog}%
  \BibitemOpen
  \bibfield  {author} {\bibinfo {author} {\bibfnamefont {M.}~\bibnamefont
  {Bordag}}, \bibinfo {author} {\bibfnamefont {G.~L.}\ \bibnamefont
  {Klimchitskaya}}, \bibinfo {author} {\bibfnamefont {V.~M.}\ \bibnamefont
  {Mostepanenko}}, \ and\ \bibinfo {author} {\bibfnamefont {V.~M.}\
  \bibnamefont {Petrov}},\ }\href {\doibase 10.1103/PhysRevD.91.045037}
  {\bibfield  {journal} {\bibinfo  {journal} {Phys. Rev. D}\ }\textbf {\bibinfo
  {volume} {91}},\ \bibinfo {pages} {045037} (\bibinfo {year}
  {2015})}\BibitemShut {NoStop}%
\bibitem [{\citenamefont {Klimchitskaya}\ \emph
  {et~al.}(2014{\natexlab{a}})\citenamefont {Klimchitskaya}, \citenamefont
  {Mohideen},\ and\ \citenamefont
  {Mostepanenko}}]{Klimchitskaya:2014:TotCifgsutptacwe}%
  \BibitemOpen
  \bibfield  {author} {\bibinfo {author} {\bibfnamefont {G.~L.}\ \bibnamefont
  {Klimchitskaya}}, \bibinfo {author} {\bibfnamefont {U.}~\bibnamefont
  {Mohideen}}, \ and\ \bibinfo {author} {\bibfnamefont {V.~M.}\ \bibnamefont
  {Mostepanenko}},\ }\href {\doibase 10.1103/PhysRevB.89.115419} {\bibfield
  {journal} {\bibinfo  {journal} {Phys. Rev. B}\ }\textbf {\bibinfo {volume}
  {89}},\ \bibinfo {pages} {115419} (\bibinfo {year}
  {2014}{\natexlab{a}})}\BibitemShut {NoStop}%
\bibitem [{\citenamefont {Klimchitskaya}\ and\ \citenamefont
  {Mostepanenko}(2014)}]{Klimchitskaya:2014:OoteitCifgs}%
  \BibitemOpen
  \bibfield  {author} {\bibinfo {author} {\bibfnamefont {G.~L.}\ \bibnamefont
  {Klimchitskaya}}\ and\ \bibinfo {author} {\bibfnamefont {V.~M.}\ \bibnamefont
  {Mostepanenko}},\ }\href {\doibase 10.1103/PhysRevA.89.052512} {\bibfield
  {journal} {\bibinfo  {journal} {Phys. Rev. A}\ }\textbf {\bibinfo {volume}
  {89}},\ \bibinfo {pages} {052512} (\bibinfo {year} {2014})}\BibitemShut
  {NoStop}%
\bibitem [{\citenamefont {Klimchitskaya}\ \emph
  {et~al.}(2014{\natexlab{b}})\citenamefont {Klimchitskaya}, \citenamefont
  {Mostepanenko},\ and\ \citenamefont
  {Sernelius}}]{Klimchitskaya:2014:TafdtCiigDcfvpt}%
  \BibitemOpen
  \bibfield  {author} {\bibinfo {author} {\bibfnamefont {G.~L.}\ \bibnamefont
  {Klimchitskaya}}, \bibinfo {author} {\bibfnamefont {V.~M.}\ \bibnamefont
  {Mostepanenko}}, \ and\ \bibinfo {author} {\bibfnamefont {B.~E.}\
  \bibnamefont {Sernelius}},\ }\href {\doibase 10.1103/PhysRevB.89.125407}
  {\bibfield  {journal} {\bibinfo  {journal} {Phys. Rev. B}\ }\textbf {\bibinfo
  {volume} {89}},\ \bibinfo {pages} {125407} (\bibinfo {year}
  {2014}{\natexlab{b}})}\BibitemShut {NoStop}%
\bibitem [{\citenamefont {Gusynin}\ \emph {et~al.}(2007)\citenamefont
  {Gusynin}, \citenamefont {Sharapov},\ and\ \citenamefont
  {Carbotte}}]{Gusynin:2007:Mcig}%
  \BibitemOpen
  \bibfield  {author} {\bibinfo {author} {\bibfnamefont {V.~P.}\ \bibnamefont
  {Gusynin}}, \bibinfo {author} {\bibfnamefont {S.~G.}\ \bibnamefont
  {Sharapov}}, \ and\ \bibinfo {author} {\bibfnamefont {J.~P.}\ \bibnamefont
  {Carbotte}},\ }\href {\doibase 10.1088/0953-8984/19/2/026222} {\bibfield
  {journal} {\bibinfo  {journal} {J. Phys.: Condens. Matter}\ }\textbf
  {\bibinfo {volume} {19}},\ \bibinfo {pages} {026222} (\bibinfo {year}
  {2007})}\BibitemShut {NoStop}%
\bibitem [{\citenamefont {Falkovsky}\ and\ \citenamefont
  {Varlamov}(2007)}]{Falkovsky:2007:Sdogc}%
  \BibitemOpen
  \bibfield  {author} {\bibinfo {author} {\bibfnamefont {L.~A.}\ \bibnamefont
  {Falkovsky}}\ and\ \bibinfo {author} {\bibfnamefont {A.~A.}\ \bibnamefont
  {Varlamov}},\ }\href {\doibase 10.1140/epjb/e2007-00142-3} {\bibfield
  {journal} {\bibinfo  {journal} {Eur. Phys J. B}\ }\textbf {\bibinfo {volume}
  {56}},\ \bibinfo {pages} {281} (\bibinfo {year} {2007})}\BibitemShut
  {NoStop}%
\bibitem [{\citenamefont {Nair}\ \emph {et~al.}(2008)\citenamefont {Nair},
  \citenamefont {Blake}, \citenamefont {Grigorenko}, \citenamefont {Novoselov},
  \citenamefont {Booth}, \citenamefont {Stauber}, \citenamefont {Peres},\ and\
  \citenamefont {Geim}}]{Nair:2008:FSCDVToG}%
  \BibitemOpen
  \bibfield  {author} {\bibinfo {author} {\bibfnamefont {R.~R.}\ \bibnamefont
  {Nair}}, \bibinfo {author} {\bibfnamefont {P.}~\bibnamefont {Blake}},
  \bibinfo {author} {\bibfnamefont {A.~N.}\ \bibnamefont {Grigorenko}},
  \bibinfo {author} {\bibfnamefont {K.~S.}\ \bibnamefont {Novoselov}}, \bibinfo
  {author} {\bibfnamefont {T.~J.}\ \bibnamefont {Booth}}, \bibinfo {author}
  {\bibfnamefont {T.}~\bibnamefont {Stauber}}, \bibinfo {author} {\bibfnamefont
  {N.~M.~R.}\ \bibnamefont {Peres}}, \ and\ \bibinfo {author} {\bibfnamefont
  {A.~K.}\ \bibnamefont {Geim}},\ }\href {\doibase 10.1126/science.1156965}
  {\bibfield  {journal} {\bibinfo  {journal} {Science}\ }\textbf {\bibinfo
  {volume} {320}},\ \bibinfo {pages} {1308} (\bibinfo {year}
  {2008})}\BibitemShut {NoStop}%
\bibitem [{\citenamefont {Khusnutdinov}\ \emph {et~al.}(2014)\citenamefont
  {Khusnutdinov}, \citenamefont {Drosdoff},\ and\ \citenamefont
  {Woods}}]{Khusnutdinov:2014:Cefswcc}%
  \BibitemOpen
  \bibfield  {author} {\bibinfo {author} {\bibfnamefont {N.}~\bibnamefont
  {Khusnutdinov}}, \bibinfo {author} {\bibfnamefont {D.}~\bibnamefont
  {Drosdoff}}, \ and\ \bibinfo {author} {\bibfnamefont {L.~M.}\ \bibnamefont
  {Woods}},\ }\href {\doibase 10.1103/PhysRevD.89.085033} {\bibfield  {journal}
  {\bibinfo  {journal} {Phys. Rev. D}\ }\textbf {\bibinfo {volume} {89}},\
  \bibinfo {pages} {085033} (\bibinfo {year} {2014})}\BibitemShut {NoStop}%
\bibitem [{\citenamefont {Djuri\v{s}i\'{c}}\ and\ \citenamefont
  {Li}(1999)}]{Djurisic:1999:Opog}%
  \BibitemOpen
  \bibfield  {author} {\bibinfo {author} {\bibfnamefont {A.}~\bibnamefont
  {Djuri\v{s}i\'{c}}}\ and\ \bibinfo {author} {\bibfnamefont {E.}~\bibnamefont
  {Li}},\ }\href {\doibase 10.1063/1.369370} {\bibfield  {journal} {\bibinfo
  {journal} {J. Appl. Phys.}\ }\textbf {\bibinfo {volume} {85}},\ \bibinfo
  {pages} {7404} (\bibinfo {year} {1999})}\BibitemShut {NoStop}%
\bibitem [{\citenamefont {Bordag}\ \emph
  {et~al.}(2009{\natexlab{b}})\citenamefont {Bordag}, \citenamefont
  {Klimchitskaya}, \citenamefont {Mohideen},\ and\ \citenamefont
  {Mostepanenko}}]{Bordag:2009:ACE}%
  \BibitemOpen
  \bibfield  {author} {\bibinfo {author} {\bibfnamefont {M.}~\bibnamefont
  {Bordag}}, \bibinfo {author} {\bibfnamefont {G.}~\bibnamefont
  {Klimchitskaya}}, \bibinfo {author} {\bibfnamefont {U.}~\bibnamefont
  {Mohideen}}, \ and\ \bibinfo {author} {\bibfnamefont {V.}~\bibnamefont
  {Mostepanenko}},\ }\href@noop {} {\emph {\bibinfo {title} {Advances in the
  Casimir Effect}}}\ (\bibinfo  {publisher} {Oxford University Press, Oxford},\
  \bibinfo {year} {2009})\ p.\ \bibinfo {pages} {749}\BibitemShut {NoStop}%
\bibitem [{\citenamefont {Khusnutdinov}\ and\ \citenamefont
  {Kashapov}(2015)}]{Khusnutdinov:2015:Cefacopcs}%
  \BibitemOpen
  \bibfield  {author} {\bibinfo {author} {\bibfnamefont {N.}~\bibnamefont
  {Khusnutdinov}}\ and\ \bibinfo {author} {\bibfnamefont {R.}~\bibnamefont
  {Kashapov}},\ }\href {\doibase 10.1007/s11232-015-0276-0} {\bibfield
  {journal} {\bibinfo  {journal} {Theor. Math. Phys.}\ }\textbf {\bibinfo
  {volume} {183}},\ \bibinfo {pages} {491} (\bibinfo {year}
  {2015})}\BibitemShut {NoStop}%
\bibitem [{\citenamefont {Marinopoulos}\ \emph {et~al.}(2004)\citenamefont
  {Marinopoulos}, \citenamefont {Reining}, \citenamefont {Rubio},\ and\
  \citenamefont {Olevano}}]{Marinopoulos:2004:isotoaawdrog}%
  \BibitemOpen
  \bibfield  {author} {\bibinfo {author} {\bibfnamefont {A.~G.}\ \bibnamefont
  {Marinopoulos}}, \bibinfo {author} {\bibfnamefont {L.}~\bibnamefont
  {Reining}}, \bibinfo {author} {\bibfnamefont {A.}~\bibnamefont {Rubio}}, \
  and\ \bibinfo {author} {\bibfnamefont {V.}~\bibnamefont {Olevano}},\ }\href
  {\doibase 10.1103/PhysRevB.69.245419} {\bibfield  {journal} {\bibinfo
  {journal} {Phys. Rev. B}\ }\textbf {\bibinfo {volume} {69}},\ \bibinfo
  {pages} {245419} (\bibinfo {year} {2004})}\BibitemShut {NoStop}%
\bibitem [{\citenamefont {Dobson}\ \emph {et~al.}(2006)\citenamefont {Dobson},
  \citenamefont {White},\ and\ \citenamefont {Rubio}}]{Dobson:2006:ADIABvdWEF}%
  \BibitemOpen
  \bibfield  {author} {\bibinfo {author} {\bibfnamefont {J.~F.}\ \bibnamefont
  {Dobson}}, \bibinfo {author} {\bibfnamefont {A.}~\bibnamefont {White}}, \
  and\ \bibinfo {author} {\bibfnamefont {A.}~\bibnamefont {Rubio}},\ }\href
  {\doibase 10.1103/PhysRevLett.96.073201} {\bibfield  {journal} {\bibinfo
  {journal} {Phys. Rev. Lett.}\ }\textbf {\bibinfo {volume} {96}},\ \bibinfo
  {pages} {073201} (\bibinfo {year} {2006})}\BibitemShut {NoStop}%
\bibitem [{\citenamefont {Dobson}\ \emph {et~al.}(2001)\citenamefont {Dobson},
  \citenamefont {McLennan}, \citenamefont {Rubio}, \citenamefont {Wang},
  \citenamefont {Gould}, \citenamefont {Le},\ and\ \citenamefont
  {Dinte}}]{Dobson:2001:PoDFITaP}%
  \BibitemOpen
  \bibfield  {author} {\bibinfo {author} {\bibfnamefont {J.~F.}\ \bibnamefont
  {Dobson}}, \bibinfo {author} {\bibfnamefont {K.}~\bibnamefont {McLennan}},
  \bibinfo {author} {\bibfnamefont {A.}~\bibnamefont {Rubio}}, \bibinfo
  {author} {\bibfnamefont {J.}~\bibnamefont {Wang}}, \bibinfo {author}
  {\bibfnamefont {T.}~\bibnamefont {Gould}}, \bibinfo {author} {\bibfnamefont
  {H.~M.}\ \bibnamefont {Le}}, \ and\ \bibinfo {author} {\bibfnamefont {B.~P.}\
  \bibnamefont {Dinte}},\ }\href {\doibase 10.1071/CH01052} {\bibfield
  {journal} {\bibinfo  {journal} {Aust. J. Chem.}\ }\textbf {\bibinfo {volume}
  {54}},\ \bibinfo {pages} {513} (\bibinfo {year} {2001})}\BibitemShut
  {NoStop}%
\bibitem [{\citenamefont {Bostr\"om}\ and\ \citenamefont
  {Sernelius}(2000)}]{Bostroem:2000:FvdWibtmf}%
  \BibitemOpen
  \bibfield  {author} {\bibinfo {author} {\bibfnamefont {M.}~\bibnamefont
  {Bostr\"om}}\ and\ \bibinfo {author} {\bibfnamefont {B.~E.}\ \bibnamefont
  {Sernelius}},\ }\href {\doibase 10.1103/PhysRevB.61.2204} {\bibfield
  {journal} {\bibinfo  {journal} {Phys. Rev. B}\ }\textbf {\bibinfo {volume}
  {61}},\ \bibinfo {pages} {2204} (\bibinfo {year} {2000})}\BibitemShut
  {NoStop}%
\bibitem [{\citenamefont {Schabel}\ and\ \citenamefont
  {Martins}(1992)}]{Schabel:1992:Eoibig}%
  \BibitemOpen
  \bibfield  {author} {\bibinfo {author} {\bibfnamefont {M.~C.}\ \bibnamefont
  {Schabel}}\ and\ \bibinfo {author} {\bibfnamefont {J.~L.}\ \bibnamefont
  {Martins}},\ }\href {\doibase 10.1103/PhysRevB.46.7185} {\bibfield  {journal}
  {\bibinfo  {journal} {Phys. Rev. B}\ }\textbf {\bibinfo {volume} {46}},\
  \bibinfo {pages} {7185} (\bibinfo {year} {1992})}\BibitemShut {NoStop}%
\bibitem [{\citenamefont {Rydberg}\ \emph {et~al.}(2003)\citenamefont
  {Rydberg}, \citenamefont {Dion}, \citenamefont {Jacobson}, \citenamefont
  {Schr{\"o}der}, \citenamefont {Hyldgaard}, \citenamefont {Simak},
  \citenamefont {Langreth},\ and\ \citenamefont
  {Lundqvist}}]{Rydberg:2003:VdWDFLS}%
  \BibitemOpen
  \bibfield  {author} {\bibinfo {author} {\bibfnamefont {H.}~\bibnamefont
  {Rydberg}}, \bibinfo {author} {\bibfnamefont {M.}~\bibnamefont {Dion}},
  \bibinfo {author} {\bibfnamefont {N.}~\bibnamefont {Jacobson}}, \bibinfo
  {author} {\bibfnamefont {E.}~\bibnamefont {Schr{\"o}der}}, \bibinfo {author}
  {\bibfnamefont {P.}~\bibnamefont {Hyldgaard}}, \bibinfo {author}
  {\bibfnamefont {S.~I.}\ \bibnamefont {Simak}}, \bibinfo {author}
  {\bibfnamefont {D.~C.}\ \bibnamefont {Langreth}}, \ and\ \bibinfo {author}
  {\bibfnamefont {B.~I.}\ \bibnamefont {Lundqvist}},\ }\href {\doibase
  10.1103/PhysRevLett.91.126402} {\bibfield  {journal} {\bibinfo  {journal}
  {Phys. Rev. Lett.}\ }\textbf {\bibinfo {volume} {91}},\ \bibinfo {pages}
  {126402} (\bibinfo {year} {2003})}\BibitemShut {NoStop}%
\bibitem [{\citenamefont {Benedict}\ \emph {et~al.}(1998)\citenamefont
  {Benedict}, \citenamefont {Chopra}, \citenamefont {Cohen}, \citenamefont
  {Zettl}, \citenamefont {Louie},\ and\ \citenamefont
  {Crespi}}]{Benedict:1998:Mdotibeig}%
  \BibitemOpen
  \bibfield  {author} {\bibinfo {author} {\bibfnamefont {L.~X.}\ \bibnamefont
  {Benedict}}, \bibinfo {author} {\bibfnamefont {N.~G.}\ \bibnamefont
  {Chopra}}, \bibinfo {author} {\bibfnamefont {M.~L.}\ \bibnamefont {Cohen}},
  \bibinfo {author} {\bibfnamefont {A.}~\bibnamefont {Zettl}}, \bibinfo
  {author} {\bibfnamefont {S.~G.}\ \bibnamefont {Louie}}, \ and\ \bibinfo
  {author} {\bibfnamefont {V.~H.}\ \bibnamefont {Crespi}},\ }\href {\doibase
  http://dx.doi.org/10.1016/S0009-2614(97)01466-8} {\bibfield  {journal}
  {\bibinfo  {journal} {Chem. Phys. Lett.}\ }\textbf {\bibinfo {volume}
  {286}},\ \bibinfo {pages} {490 } (\bibinfo {year} {1998})}\BibitemShut
  {NoStop}%
\bibitem [{\citenamefont {Zacharia}\ \emph {et~al.}(2004)\citenamefont
  {Zacharia}, \citenamefont {Ulbricht},\ and\ \citenamefont
  {Hertel}}]{Zacharia:2004:Iceogftdoph}%
  \BibitemOpen
  \bibfield  {author} {\bibinfo {author} {\bibfnamefont {R.}~\bibnamefont
  {Zacharia}}, \bibinfo {author} {\bibfnamefont {H.}~\bibnamefont {Ulbricht}},
  \ and\ \bibinfo {author} {\bibfnamefont {T.}~\bibnamefont {Hertel}},\ }\href
  {\doibase 10.1103/PhysRevB.69.155406} {\bibfield  {journal} {\bibinfo
  {journal} {Phys. Rev. B}\ }\textbf {\bibinfo {volume} {69}},\ \bibinfo
  {pages} {155406} (\bibinfo {year} {2004})}\BibitemShut {NoStop}%
\end{thebibliography}
